\documentclass[12pt]{article}
\usepackage{graphicx}

\def\vb#1{\vbox to #1 pt{}}
\def\Slash#1{#1\!\!\!\! /}
\def\ifmath#1{\relax\ifmmode #1\else $#1$\fi}
\def\half{\ifmath{{\textstyle{1 \over 2}}}}
\def\nn{\nonumber}
\def\ds{\displaystyle}
\def\ra{\rightarrow}
\def\rp{$R_p \hspace{-1em}/\;\:$ }
\def\bold#1{\setbox0=\hbox{$#1$} 
      \kern-.025em\copy0\kern-\wd0 
      \kern.05em\copy0\kern-\wd0 
      \kern-.025em\raise.0433em\box0 }
\def\Eq#1{{Eq. (\ref{#1})}}
\newcommand {\chiz} [1] {\tilde{\chi}^{0}_{#1} }

\begin{document}
\begin{flushright}
FISIST/06--2002/CFIF\\
hep-ph/0202149 
\end{flushright} 

\begin{center}  

{\Large \bf Neutrino Masses and Mixings in SUSY with Broken R-Parity}
{\footnote{To appear in the Proceedings of the Corfu Summer
Institute, Corfu 2001, 
Edited by J. Rizos.}}

\vspace{5mm}

{\large M. Hirsch${}^{a,b}$,
W. Porod${}^{a,c}$, M.A. D\'{\i}az${}^d$, 
\underline{J.C. Rom\~ao}${}^e$ and 
J.W.F. Valle${}^a$ } 

\vspace{4mm}

{\small \it $^a$ Departament de F\'\i sica Te\`orica, IFIC--CSIC,
          Universitat de Val\`encia\\
          46100 Burjassot, Val\`encia, Spain \\ }
{\small \it $^b$ Department of Physics and Astronomy, 
          University of Southampton\\ Southampton, SO17 1BJ, U.K.\\}
{\small \it $^c$ Inst.~f\"ur Theor. Physik, Universit\"at Z\"urich,
          CH-8057 Z\"urich, 
          Switzerland\\}
{\small \it $^d$ Facultad de F\'\i sica, Universidad Cat\'olica de Chile\\ 
          Av. Vicu\~na Mackenna 4860, Santiago, Chile\\}
{\small \it $^e$ Departamento de F\'\i sica, Instituto Superior T\'ecnico\\ 
          Av. Rovisco Pais 1, 1049-001 Lisboa, Portugal \\}

\end{center}

\begin{abstract}
\noindent
The simplest unified extension of the Minimal Supersymmetric Standard
Model with bilinear R--Parity violation provides a predictive scheme
for neutrino masses which can account for the observed atmospheric and
solar neutrino anomalies.
Despite the smallness of neutrino masses R-parity violation is
observable at present and future high-energy colliders, providing an
unambiguous cross-check of the model.
\end{abstract}

\section{Introduction}
The announcement of high statistics atmospheric neutrino data
by the SuperKamio\-kande collaboration \cite{kamiokande} has
confirmed the deficit of muon neutrinos, especially at small zenith
angles, opening a new era in neutrino physics. 
Although there may be alternative solutions of the atmospheric
neutrino anomaly ~\cite{Gonzalez-Garcia} it is fair to say that
the simplest interpretation of the data is in terms of $\nu_{\mu}$ to
$\nu_{\tau}$ 
flavor oscillations with maximal mixing. This excludes a large mixing
among $\nu_{\tau}$ and $\nu_e$~\cite{kamiokande}, in agreement also
with the CHOOZ reactor data~\cite{chooz}.   On the other hand the
persistent disagreement between solar neutrino data and theoretical
expectations~\cite{BP98} has been a long-standing problem in physics. 
Recent solar neutrino
data~\cite{Smy:2001yn} are consistent with both
vacuum oscillations and MSW conversions. In the latter case one can
have either the small or the large mixing angle solutions, with 
the latter being clearly preferred ~\cite{Gonzalez-Garcia}.

Many attempts have appeared in the literature to explain the
data. Here we review recent  results~\cite{numass} obtained in a 
model~\cite{epsmodel} which is a simple
extension of the MSSM with bilinear R-parity violation (BRpV). 
This model, despite being a minimal extension of the MSSM, can explain
the solar and atmospheric neutrino data. Its most attractive feature
is that it gives definite predictions for accelerator physics for the
same range of parameters that explain the neutrino data.

\section{Bilinear R-Parity Violation (BRpV)}
\subsection{The Model}

The superpotential $W$ is given by 
\begin{equation}
W
\hskip -1mm=\hskip -0.5mm
\varepsilon_{ab}\! \left[
 h_U^{ij}\widehat Q_i^a\widehat U_j\widehat H_u^b
\!+h_D^{ij}\widehat Q_i^b\widehat D_j\widehat H_d^a
\!+h_E^{ij}\widehat L_i^b\widehat R_j\widehat H_d^a 
\! -\mu\widehat H_d^a\widehat H_u^b
\!+\epsilon_i\widehat L_i^a\widehat H_u^b\right]
\end{equation}
while the set of soft supersymmetry
breaking terms are
\begin{eqnarray}
V_{soft}&\hskip -5mm=\hskip -5mm&
M_Q^{ij2}\widetilde Q^{a*}_i\widetilde Q^a_j+M_U^{ij2}
\widetilde U^*_i\widetilde U_j+M_D^{ij2}\widetilde D^*_i
\widetilde D_j+M_L^{ij2}\widetilde L^{a*}_i\widetilde L^a_j
+M_R^{ij2}\widetilde R^*_i\widetilde R_j
\cr
\vb{18}
&&\hskip -5mm
+m_{H_d}^2 H^{a*}_d H^a_d+m_{H_u}^2 H^{a*}_u H^a_u 
- \left[\sum_i \half M_i\lambda_i\lambda_i
+\varepsilon_{ab}\left(
A_U^{ij}\widetilde Q^a_i\widetilde U_j H_u^b \right. \right. \cr
&&\hskip -5mm
\left. \vb{18} \left.
+A_D^{ij}\widetilde Q^b_i\widetilde D_j H_d^a
+A_E^{ij}\widetilde L^b_i\widetilde R_j H_d^a 
-B\mu H_d^a H_u^b+B_i\epsilon_i\widetilde L^a_i H_u^b \right) 
+ h.c \right]
\end{eqnarray}

\noindent
The bilinear
R-parity violating term \textit{cannot} be eliminated by superfield
redefinition.
The reason is \cite{marco} 
that the bottom Yukawa coupling, usually neglected,
plays a crucial role in splitting
the soft-breaking parameters $B$ and $B_i$ as well as the scalar
masses $m_{H_d}^2$ and $M_L^{2}$, assumed to be equal at the
unification scale.
The BRpV model is a 1(3) parameter(s) generalization of the MSSM.
It can be thought as an \textit{effective} model
showing the more important features of the Spontaneous Broken R-parity
model (SBRP)\cite{OriginalSBRP,SBRpV} at the weak scale. 
The mass matrices, charged and neutral currents, are similar to the
SBRP--model if we identify
\begin{equation}
\epsilon \equiv v_R h_{\nu}
\end{equation}
The model has the MSSM as a limit when $\epsilon_i\ra 0$.

\subsection{Radiative Breaking}

At $Q = M_{GUT}$ we assume the standard minimal supergravity
unifications assumptions, 
\begin{eqnarray}
&&A_t = A_b = A_{\tau} \equiv A \:, B=B_2=A-1 \:, \cr
\vb{18}
&&
m_{H_d}^2 = m_{H_u}^2 = M_{L}^2 = M_{R}^2 = 
M_{Q}^2 =M_{U}^2 = M_{D}^2 = m_0^2 \:,  \cr
\vb{18}
&&
M_3 = M_2 = M_1 = M_{1/2} 
\end{eqnarray}
In order to determine the values of the Yukawa couplings and of the
soft breaking scalar masses at low energies we first run the RGE's from
the unification scale $M_{GUT} \sim 10^{16}$ GeV down to the weak
scale. For details see~\cite{numass,epsmodel}.

\section{Tree Level Neutrino Masses and Mixings}

\subsection{Neutral fermion mass matrix}

In the basis $\psi^{0T}= 
(-i\lambda',-i\lambda^3,\widetilde{H}_d^1,\widetilde{H}_u^2,
\nu_{e}, \nu_{\mu}, \nu_{\tau} )$ 
the neutral fer\-mions mass terms in the Lagrangian are given by 
\begin{equation}
{\cal L}_m=-\frac 12(\psi^0)^T{\bold M}_N\psi^0+h.c.   
\end{equation}
where the neutralino/neutrino mass matrix is 
\begin{equation}
{\bold M}_N=\left[  
\begin{array}{cc}  
{\cal M}_{\chi^0}& m^T \cr
m & 0 \cr
\end{array}
\right]
\end{equation}
with
\begin{equation}
{\cal M}_{\chi^0}=\left[  
\begin{array}{cccc}  
M_1 & 0 & -\frac 12g^{\prime }v_d & \frac 12g^{\prime }v_u \cr
0 & M_2 & \frac 12gv_d & -\frac 12gv_u \cr
-\frac 12g^{\prime }v_d & \frac 12gv_d & 0 & -\mu  \cr
\frac 12g^{\prime }v_u & -\frac 12gv_u &  -\mu & 0  \cr
\end{array}  
\right] 
\quad ; \quad
m=\left[  
\begin{array}{c}
a_1 \cr
a_2 \cr
a_3 
\end{array}  
\right] 
\end{equation}
where $a_i=(-\frac 12g^{\prime }v_i, \frac 12gv_i, 0,\epsilon_i)$. 
This neutralino/neutrino mass matrix is diagonalized by 
\begin{equation}
{\cal  N}^*{\bold M}_N{\cal N}^{-1}={\rm diag}(m_{\chi^0_1},m_{\chi^0_2}, 
m_{\chi^0_3},m_{\chi^0_4},m_{\nu_1},m_{\nu_2},m_{\nu_3}) 
\label{eq:NeuMdiag} 
\end{equation}

\subsection{Approximate diagonalization of mass matrices}

If the \rp parameters are small, then
\begin{equation}
\xi = m \cdot {\cal M}_{\chi^0}^{-1}\hskip 10mm
\Rightarrow\hskip 10mm \forall \xi_{ij} \ll 1
\end{equation}
one can find an approximate solution for the mixing  matrix ${\cal N}$.
Explicit expressions can be found in Ref.~\cite{numass}.
In leading order in $\xi$ the mixing matrix ${\cal N}$ is given by,
\begin{equation}
\label{mixing}
{\cal N}^*  =  \left(\begin{array}{cc}
N^* & 0\\
0& V_\nu^T \end{array}\right)
\left(\begin{array}{cc}
1 -{1 \over 2} \xi^{\dagger} \xi& \xi^{\dagger} \\
-\xi &  1 -{1 \over 2} \xi \xi^\dagger
\end{array}\right) 
\end{equation}
This decomposition  block--diagonalizes 
${\bold M}_N$ approximately to the form 
diag($m_{eff}, {\cal M}_{\chi^0}$), where
\begin{equation}
m_{eff} = - m \cdot {\cal M}_{\chi^0}^{-1} m^T = 
\frac{M_1 g^2 + M_2 {g'}^2}{4\, det({\cal M}_{\chi^0})} 
\left(\begin{array}{ccc}
\Lambda_e^2 & \Lambda_e \Lambda_\mu
& \Lambda_e \Lambda_\tau \\
\Lambda_e \Lambda_\mu & \Lambda_\mu^2
& \Lambda_\mu \Lambda_\tau \\
\Lambda_e \Lambda_\tau & \Lambda_\mu \Lambda_\tau & \Lambda_\tau^2
\end{array}\right).
\end{equation}
The submatrices $N$ and $V_{\nu}$ in \Eq{mixing}  diagonalize 
${\cal M}_{\chi^0}$ and $m_{eff}$ 
\begin{equation}
N^{*}{\cal M}_{\chi^0} N^{\dagger} = diag(m_{\chi^0_i})
\quad ; \quad
V_{\nu}^T m_{eff} V_{\nu} = diag(0,0,m_{\nu}),
\end{equation}
where 
\begin{equation}
m_{\nu} = Tr(m_{eff}) = 
\frac{M_1 g^2 + M_2 {g'}^2}{4\, det({\cal M}_{\chi^0})} 
|{\vec \Lambda}|^2.
\end{equation}

\section{One Loop Neutrino Masses and Mixings}
\subsection{Definition}

The Self--Energy for the neutralino/neutrino is

\begin{equation}
\hskip 3.5cm \equiv
i \left\{ \slash{p} \left[ P_L \Sigma^L_{ij} + P_R \Sigma^R_{ij} \right]
-\left[ P_L \Pi^L_{ij} + P_R \Pi^R_{ij} \right]\right\}
\end{equation}

\begin{picture}(0,0)
\put(-0.5,-0.25){\includegraphics[width=3.5cm]{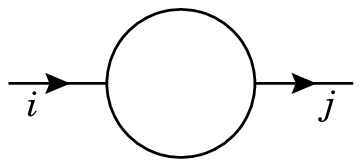}}
\end{picture}

\noindent
Then 
\begin{equation}
M^{\rm pole}_{ij}= M^{\rm \overline{DR}}_{ij}(\mu_R) + \Delta M_{ij}
\end{equation}
with
\begin{equation}
\Delta M_{ij}\! =\! \left[ \half 
\left(\Pi^V_{ij}(m_i^2)\! +\! \Pi^V_{ij}(m_j^2)\right) 
\!-\! \half 
\left( m_{\chi^0_i} \Sigma^V_{ij}(m_i^2) \! +\! 
m_{\chi^0_j} \Sigma^V_{ij}(m_j^2) \right) \right]_{\Delta=0}
\end{equation}
where
\begin{equation}
\Sigma^V=\half \left(\Sigma^L+\Sigma^R\right)
\quad ; \quad
\Pi^V=\half \left(\Pi^L+\Pi^R\right)
\end{equation}
and
\begin{equation}
\ds \Delta=\frac{2}{4-d} -\gamma_E + \ln 4\pi
\end{equation}

\subsection{Diagrams Contributing}

In a generic way the diagrams contributing are

\begin{tabular}{cccc}
\begin{picture}(0,3)
\put(-0.2,0){\includegraphics[width=30mm]{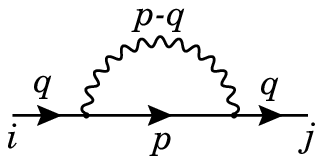}}
\end{picture}&
\begin{picture}(0,0)
\put(2.65,0){\includegraphics[width=30mm]{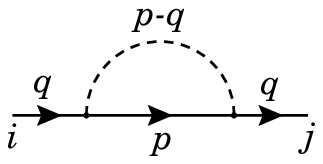}}
\end{picture}&
\begin{picture}(0,0)
\put(5.5,0){\includegraphics[width=30mm]{sself.eps}}
\end{picture}&
\begin{picture}(0,0)
\put(8.35,0.2){\includegraphics[width=30mm]{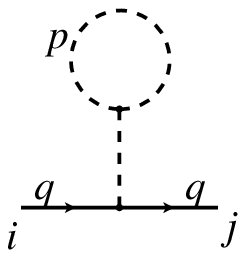}}
\end{picture}
\end{tabular}

\noindent
These diagrams can be calculated in a straightforward way. For
instance the $W$ diagram in the $\xi=1$ gauge gives
\begin{eqnarray}
\Sigma^V_{ij}&\!=\!\!& -\frac{1}{16\pi^2}\, \sum_{k=1}^5
2 \left(O^{\rm ncw}_{L jk} O^{\rm cnw}_{L ki} +
O^{\rm ncw}_{R jk} O^{\rm cnw}_{R ki}\right) B_1(p^2,m^2_k,m^2_W)\cr
\vb{35}
\Pi^V_{ij}&\!=\!\!& -\frac{1}{16\pi^2}\, \sum_{k=1}^5
(-4) \left(O^{\rm ncw}_{L jk} O^{\rm cnw}_{R ki} +
O^{\rm ncw}_{R jk} O^{\rm cnw}_{L ki}\right) m_k\, B_0(p^2,m^2_k,m^2_W)
\nn
\end{eqnarray}
where $B_0$ and $B_1$ are the Passarino-Veltman functions, and
$O^{\rm cnw}$, $O^{\rm ncw}$ are coupling matrices. Explicit
expressions can be found in~\cite{numass}.

\subsection{Gauge Invariance}

When calculating the self--energies the question of gauge invariance
arises. We have performed a careful calculation in an arbitrary
$R_{\xi}$ gauge and showed~\cite{numass} that the result was independent of the
gauge parameter $\xi$.

\subsection{The One--Loop Mass Matrix}

The one--loop corrected mass matrix is
\begin{equation}
M^{1L}= M^{0L}_{diag} + \Delta M^{1L}
\end{equation}
where
\begin{equation}
M^{0L}_{diag}= {\cal N} M_N {\cal N}^T
\end{equation}
Now we diagonalize the 1--loop mass matrix 
\begin{equation}
M^{1L}_{diag}={\cal N'} M^{1L} {\cal N'}^T
\end{equation}
Then the mass eigenstates are related to the weak basis states by
\begin{equation}
\chi_0^{mass}= {\cal N}^{1L}_{i \alpha}\, \chi_0^{weak}
\end{equation}
with
\begin{equation}
{\cal N}^{1L} = {\cal N'}\ {\cal N}
\end{equation}
The usual convention in neutrino physics
\begin{equation}
\nu_{\alpha} = U_{\alpha k}\,  \nu_k
\end{equation}
is recovered in our notation as
\begin{equation}
U_{\alpha k}= {\cal N}^{1L}_{4+k, 4+\alpha} 
\end{equation}

\subsection{Approximate Formulas for Masses and Mixings}

\subsubsection{The masses}

Looking at the numerical results~\cite{numass} we found that the most
important contribution came from the bottom-sbottom loop. To gain an
analytical understanding of the results we expanded the exact results
in the small $\Slash{R}_P$ parameters. The result is~\cite{HPVR2002}
\begin{equation}
  \label{eq:approx_mass}
M_{\nu}\simeq c_0 \left(\matrix{
\Lambda_1 \Lambda_1&\Lambda_1 \Lambda_2&\Lambda_1 \Lambda_3\cr
\Lambda_2 \Lambda_1&\Lambda_2 \Lambda_2&\Lambda_2 \Lambda_3\cr
\Lambda_3 \Lambda_1&\Lambda_3 \Lambda_2&\Lambda_3 \Lambda_3}\right)
+c_1 \left(\matrix{
\epsilon_1 \epsilon_1&\epsilon_1 \epsilon_2&\epsilon_1 \epsilon_3\cr
\epsilon_2 \epsilon_1&\epsilon_2 \epsilon_2&\epsilon_2 \epsilon_3\cr
\epsilon_3 \epsilon_1&\epsilon_3 \epsilon_2&\epsilon_3 \epsilon_3}\right)
\end{equation}
where
\begin{equation}
c_0=\frac{M_1\, g^2 +M_2\, g'^2}{4\, \det(M_{\chi^0})}
\quad ; \quad
c_1=\frac{3}{16\pi^2}\, m_b\, \sin 2 \theta_b \, \frac{h^2_b}{\mu^2}\,
\log \frac{m^2_{\tilde{b_2}}}{m^2_{\tilde{b_1}}}
\end{equation}

\begin{figure}[htb]
\begin{tabular}{cc}
\begin{picture}(7.0,5)
\put(0,0){\includegraphics[width=7cm]{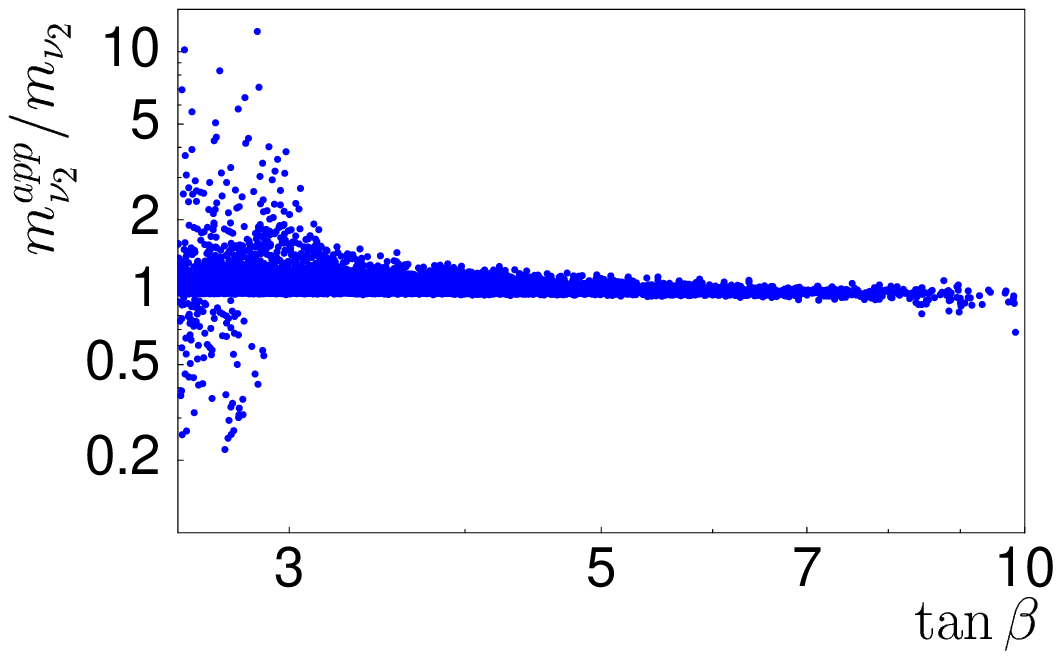}}
\end{picture}
&
\begin{picture}(7.0,5)
\put(0,0){\includegraphics[width=7cm]{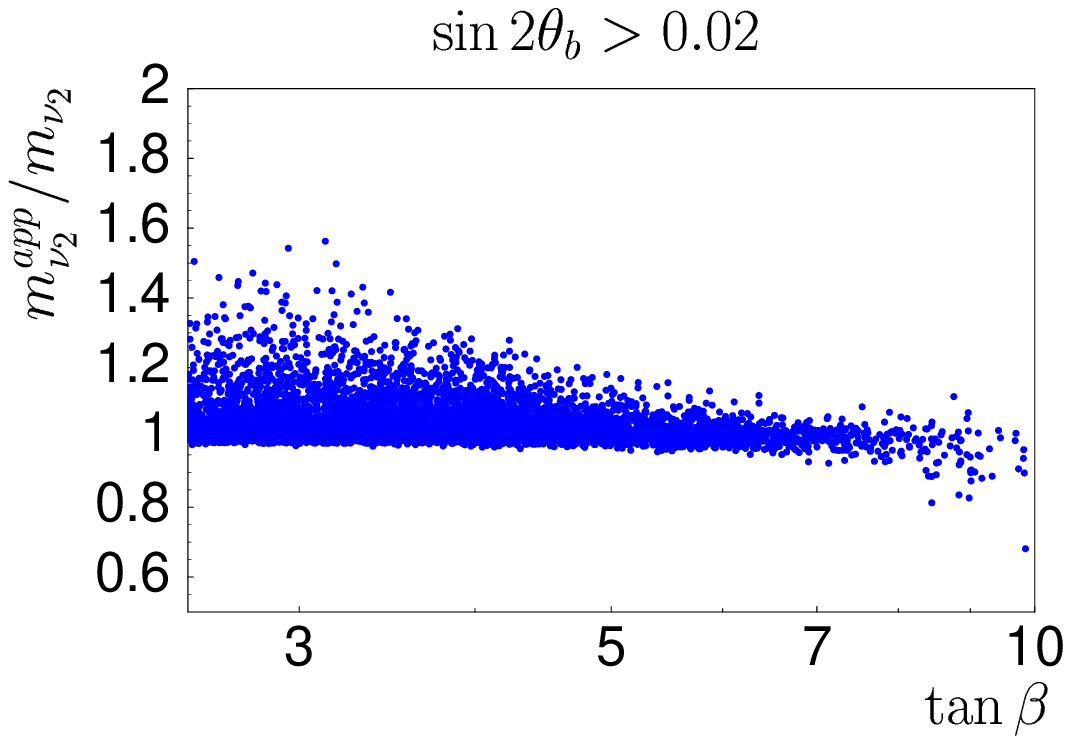}}
\end{picture}
\end{tabular}
\vspace{-3mm}
\caption{\small
a) Ratio of $m^{app}_{\nu_2}$, given by \Eq{mneu2}, with the
  exact numerical result. b) The same with the cut $sin 2 \theta_b > 0.02$.
}
\label{corfu_fig1}
\end{figure}

\noindent
Diagonalization of the mass matrix gives~\cite{HPVR2002}
\begin{eqnarray}
\label{mneu1}
m_{\nu_1}&=& 0 \\[-5mm]
\label{mneu2}
m_{\nu_2}&\simeq& \frac{3}{16\pi^2}\, m_b\, \sin 2 \theta_b \,
\frac{h^2_b}{\mu^2} \ \log \frac{m^2_{\tilde{b_2}}}{m^2_{\tilde{b_1}}}\
 \frac{(\vec \epsilon \times \vec
  \Lambda)^2}{|\vec \Lambda|^2}\\[-5mm]
\label{mneu3}
m_{\nu_3}&\simeq& \frac{M_1\, g^2 +M_2\, g'^2}{4\, \det(M_{\chi^0})}\
|\vec \Lambda|^2
\end{eqnarray}
The formula for $m_{\nu_3}$ is the tree-level formula that we used to
fix the scale of the atmospheric neutrinos by choosing $|\vec
\Lambda|$.
The quality of the approximation formula for $m_{\nu_2}$ can be seen
in Fig.~\ref{corfu_fig1}. Only for small $\tan \beta$ and very
small $b,\tilde{b}$ mixing the approximation can lead, in some cases,
to incorrect results.
Details of the derivation  can
be found in Ref.~\cite{HPVR2002} where the second most important
contribution, coming from the loop with charged Higgs/charged leptons, is also
discussed.

\subsubsection{The mixings}

The atmospheric angle is easily obtained in terms of the ratio
$\Lambda_2/\Lambda_3$. For the solar angle in the same approximation
we also get a simple formula~\cite{HPVR2002},
\begin{equation}
\tan^2 \theta_{sol} = \frac{2 \epsilon^2}{(\epsilon_2+\epsilon_3)^2}
\end{equation}
that is also in very good agreement with the exact result.

\section{Results for the Solar and Atmospheric Neutrinos}

\subsection{The masses}

The BRpV model produces a hierarchical mass spectrum for almost all
choices of parameters. The largest mass can be estimated by the tree
level value using Eq.~(\ref{mneu3}). 
Correct $\Delta m^2_{atm}$ can be easily obtained by an appropriate
choice of $| \vec \Lambda|$. The mass scale for the solar neutrinos is
generated at 1--loop level and therefore depends in a complicated way
in the model parameters. However, in most cases the result of
Eq.~(\ref{mneu2}) is a good approximation and there is no problem in
having both $\Delta m^2_{atm}$ and $\Delta m^2_{solar}$ set to the
correct scales.

\subsection{The mixings}

\begin{figure}[htb]
\begin{tabular}{cc}
\begin{picture}(7.0,6.75)
\put(0,0){\includegraphics[width=7cm]{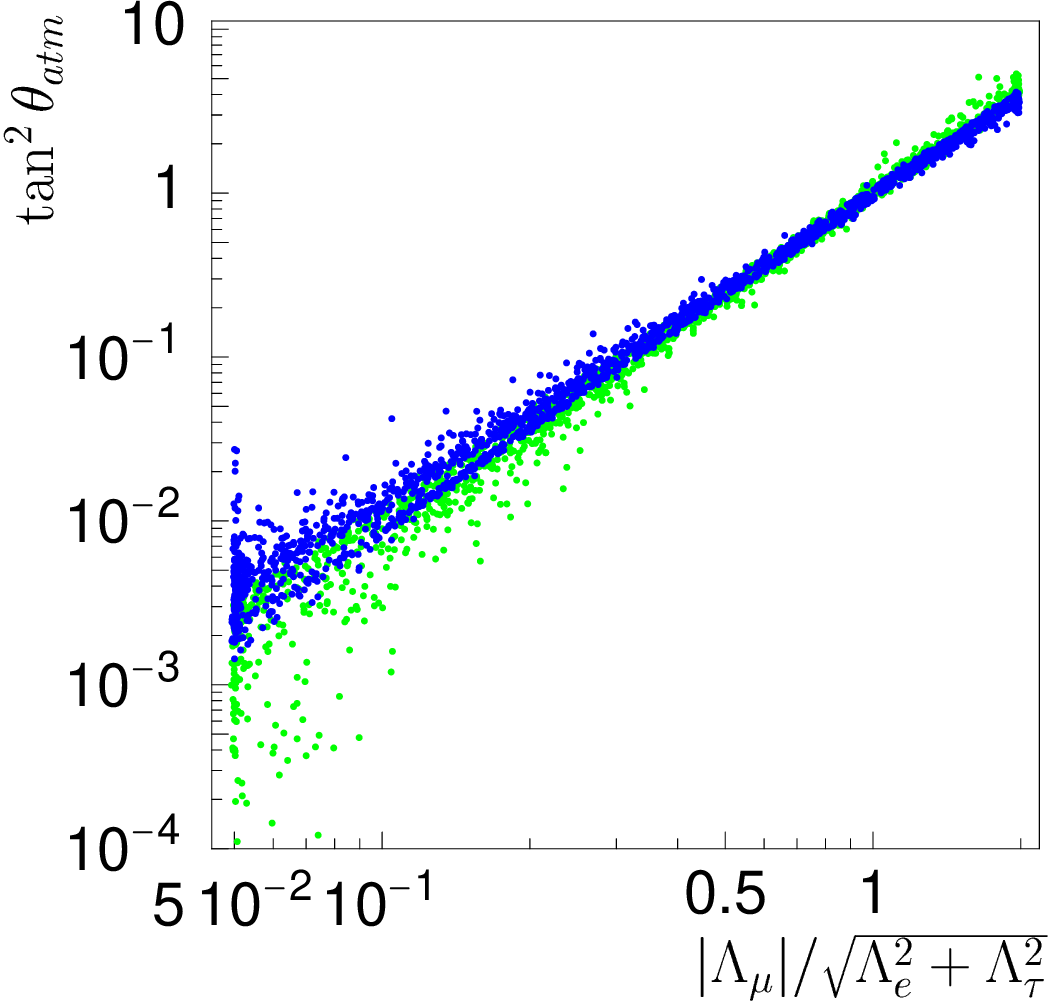}}
\end{picture}
&
\begin{picture}(7.0,6.75)
\put(0,0){\includegraphics[width=7cm]{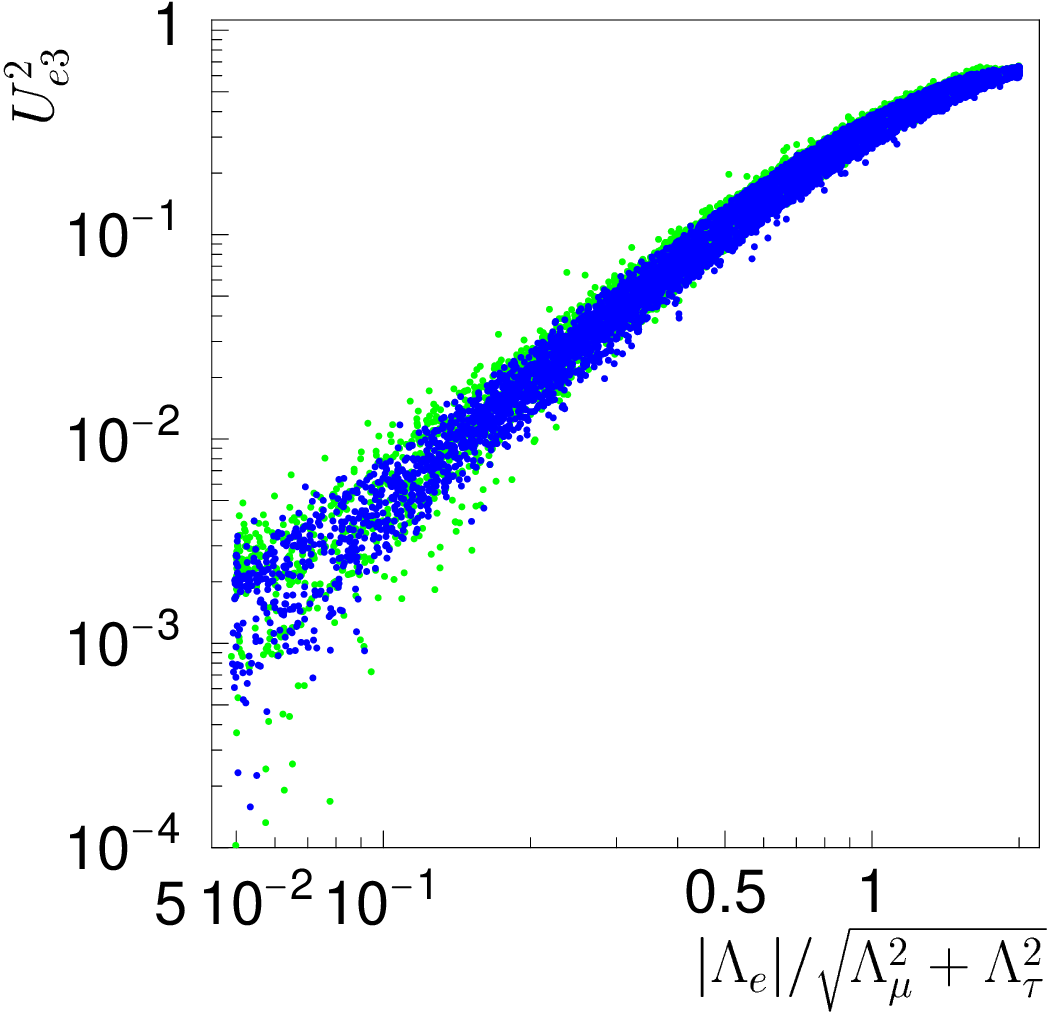}}
\end{picture}
\end{tabular}
\vspace{-3mm}
\caption{\small 
a) Atmospheric angle as a function of
  $|\Lambda_{\mu}|/\sqrt{\Lambda^2_{e}+\Lambda^2_{\tau}}$. 
  b) $U^2_{e3}$ as a function of
  $|\Lambda_e|/\sqrt{\Lambda^2_{\mu}+\Lambda^2_{\tau}}$.  
}
\label{corfu_fig2}
\end{figure}

Now we turn to the discussion of the mixing angles. We have found that if 
$\epsilon^2/|\vec \Lambda| \ll 100$ then the 1--loop corrections are
not larger than the tree level results and the flavor composition of
the 3rd mass eigenstate is approximately given by
\begin{equation}
U_{\alpha 3}\approx\Lambda_{\alpha}/|\vec \Lambda |
\end{equation}
As the atmospheric and reactor neutrino data tell us that
$\nu_{\mu}\ra \nu_{\tau}$ oscillations are preferred over 
$\nu_{\mu}\ra \nu_e$, we conclude that  
\begin{equation}
\Lambda_e \ll \Lambda_{\mu} \simeq \Lambda_{\tau}
\end{equation}
are required for BRpV to fit the data. This is sown in
Fig.~\ref{corfu_fig2} a). We cannot get so easily maximal mixing for
solar neutrinos, because in this case $U_{e 3}$ would be too large
contradicting the CHOOZ result as shown in Fig.~\ref{corfu_fig2} b).

\begin{figure}[htb]
\begin{tabular}{cc}
\begin{picture}(7.0,6.5)
\put(0,-0.2){\includegraphics[width=7cm,height=6.2cm]{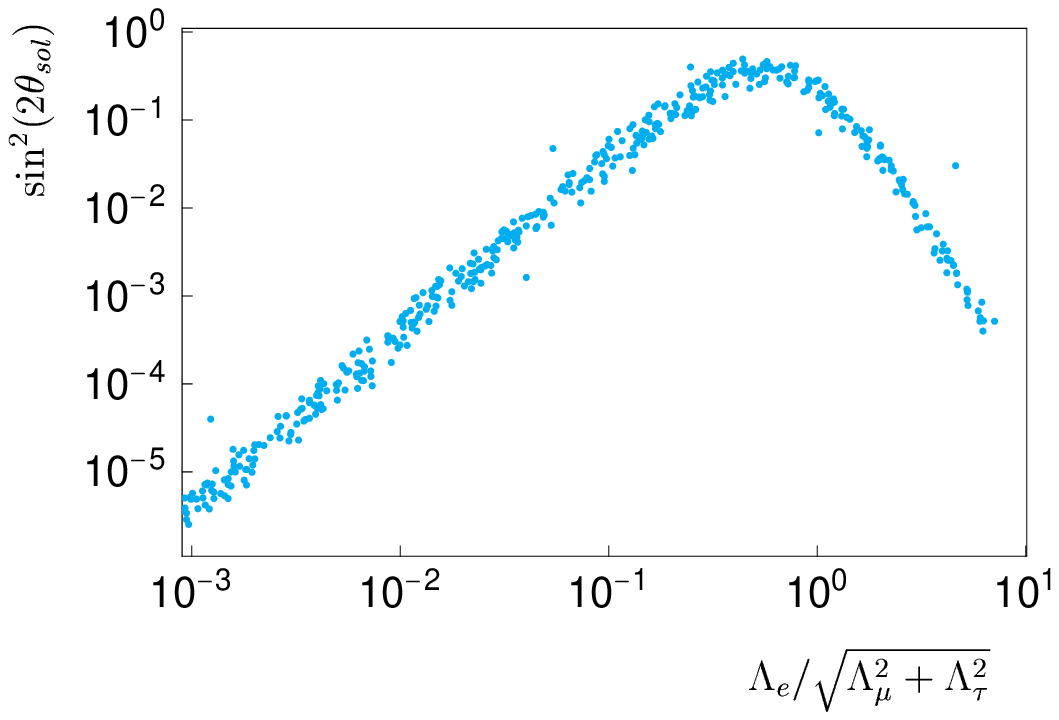}}
\end{picture}
&
\begin{picture}(7.0,6.5)
\put(0,0){\includegraphics[width=7cm,height=6cm]{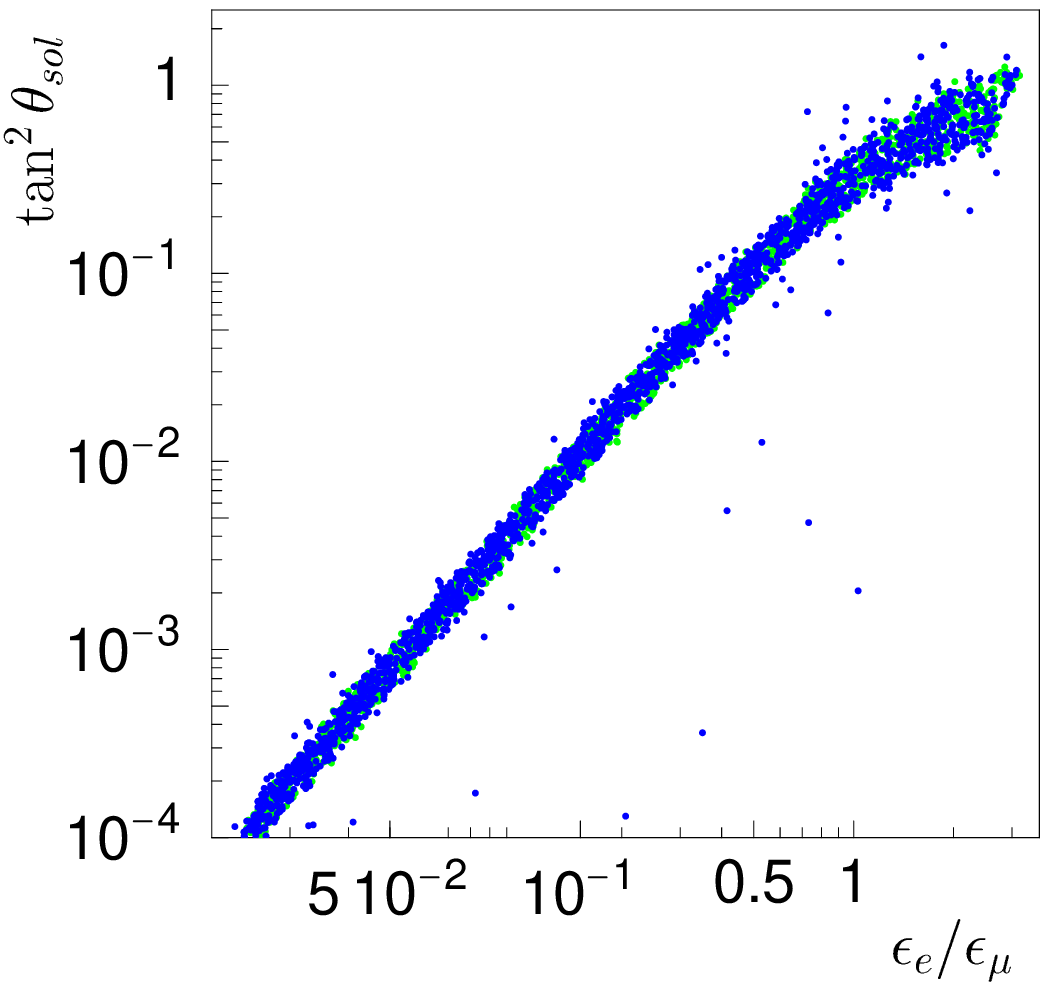}}
\end{picture}
\end{tabular}
\vspace{-3mm}
\caption{\small
Solar angle as function of: a)
$|\Lambda_e|/\sqrt{\Lambda^2_{\mu}+\Lambda^2_{\tau}}$ \ ;  
b) $\epsilon_e/\epsilon_{\mu}$.
}
\label{corfu_fig3}
\end{figure}

\noindent
We have then two scenarios. In the first one, that we call the
\textit{mSUGRA} case, we have universal boundary conditions of the soft
SUSY breaking terms. In this case we can show~\cite{numass} that 
\begin{equation}
  \label{eq:msugra}
\frac{\epsilon_e}{\epsilon_{\mu}}\simeq \frac{\Lambda_e}{\Lambda_{\mu}}  
\end{equation}
Then from Fig.~\ref{corfu_fig2} b) and the CHOOZ constraint on
$U^2_{e3}$, we conclude that \textit{both} ratios in
Eq.~(\ref{eq:msugra}) have to be small. Then from
Fig.~\ref{corfu_fig3} we conclude that the only possibility is
the small angle mixing solution for the solar neutrino problem. In the
second scenario, which we call the \textit{MSSM} case, we consider
non--universal boundary conditions of the soft SUSY breaking terms. We
have shown that even a very small deviation from universality of the soft
parameters at the GUT scale relaxes this constraint. 
In this case 
\begin{equation}
\frac{\epsilon_e}{\epsilon_{\mu}}\not=\frac{\Lambda_e}{\Lambda_{\mu}}  
\end{equation}
Then we can have at the same time \textbf{small} $U_{e3}^2$ determined by
$\Lambda_e/\Lambda_{\mu}$ as in Fig.~\ref{corfu_fig2} b) 
and \textbf{large} $\tan^2(\theta_{solar})$ determined by
$\epsilon_e/\epsilon_{\mu}$ as in Fig.~\ref{corfu_fig3} b).

\section{Probing Neutrino Mixing via Neutralino Decays}

\begin{figure}[htb]
\begin{tabular}{cc}
\begin{picture}(7.0,6.5)
\put(0,0){\includegraphics[width=7cm,height=6.5cm]{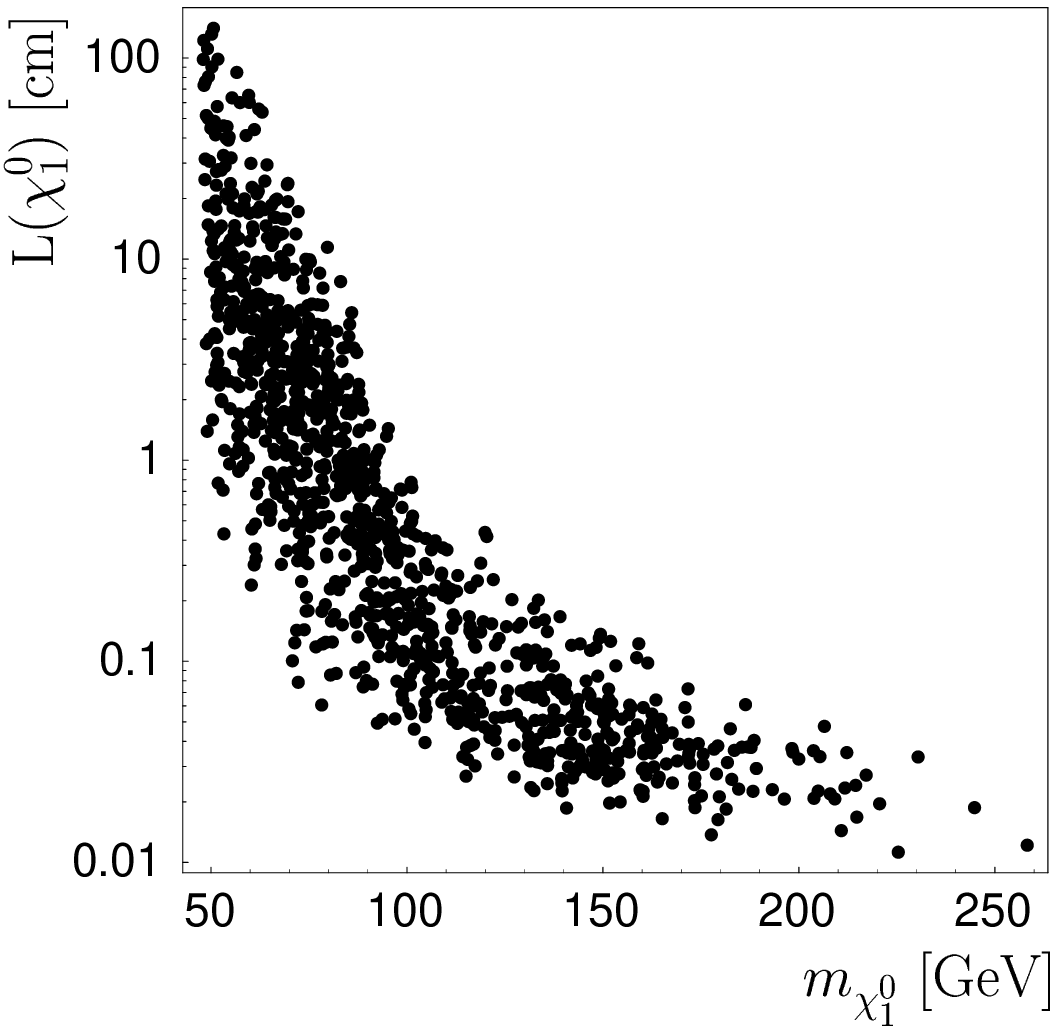}}
\end{picture}
&
\begin{picture}(7.0,6.6)
\put(0,0){\includegraphics[width=7cm,height=6.6cm]{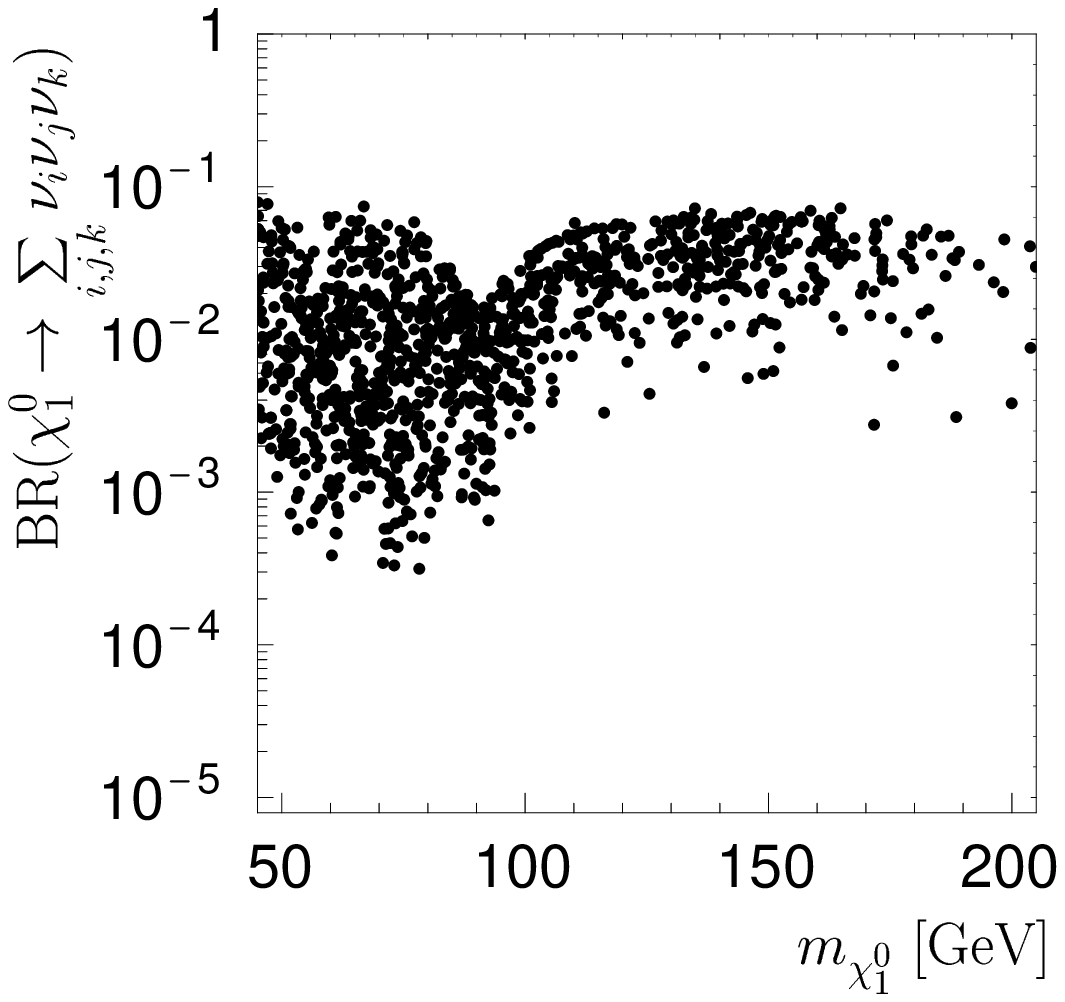}}
\end{picture}
\end{tabular}
\vspace{-3mm}
\caption{\small
a) Decay length of $\chiz{1}$ as function of $m_{\chiz{1}}$.\, 
b) Invisible branching ratio of the $\chiz{1}$ as function of
$m_{\chiz{1}}$ 
}
\label{corfu_fig4}
\end{figure}

\begin{figure}[htb]
\begin{tabular}{cc}
\begin{picture}(7,6.5)
\put(0,0){\includegraphics[width=7cm,height=6.5cm]{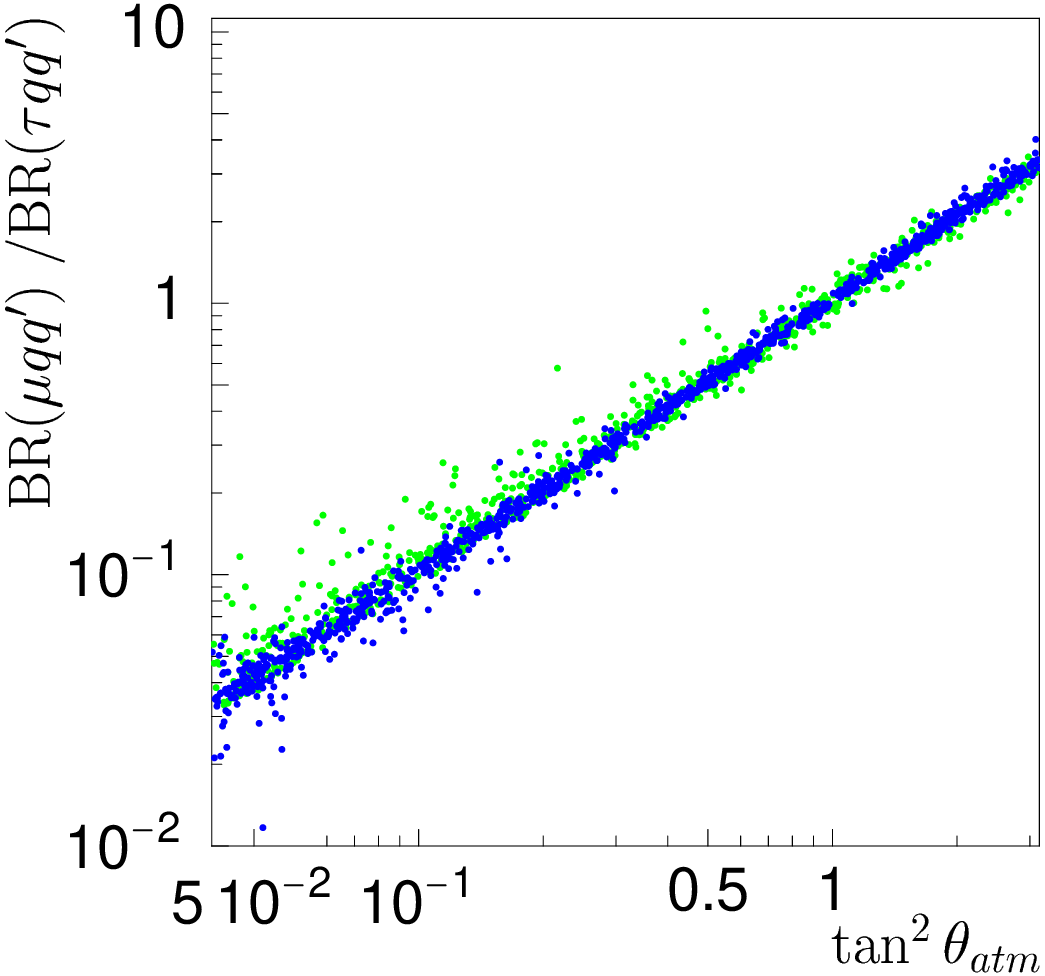}}
\end{picture}
&
\begin{picture}(7,6.5)
\put(0,-0.1){\includegraphics[width=7cm,height=6.5cm]{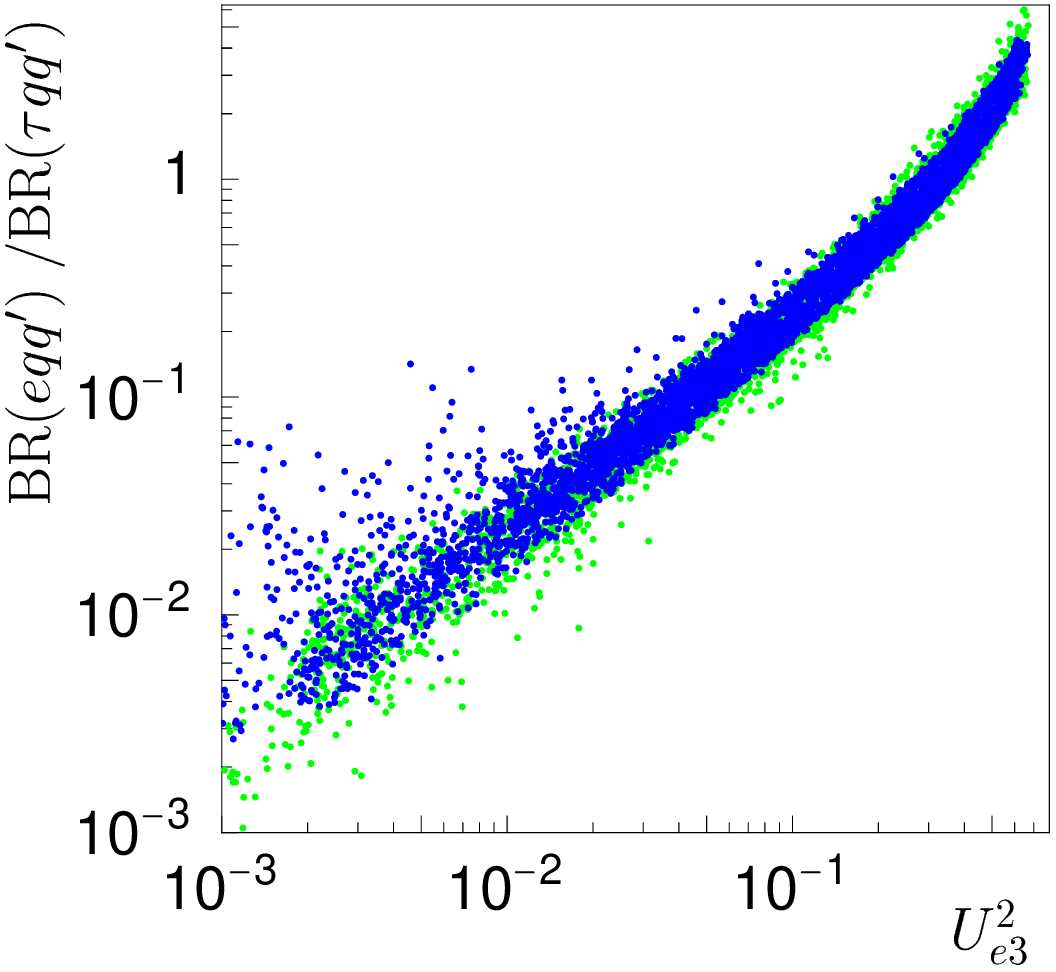}}
\end{picture}
\end{tabular}
\vspace{-3mm}
\caption{\small
Ratios of semileptonic branching ratios as functions of $\tan
\theta_{atm}$ and $U^2_{e3}$.
}

\label{corfu_fig5}
\end{figure}

\begin{figure}[htb]
\begin{tabular}{cc}
\begin{picture}(7.0,6.5)
\put(0,0){\includegraphics[width=7cm]{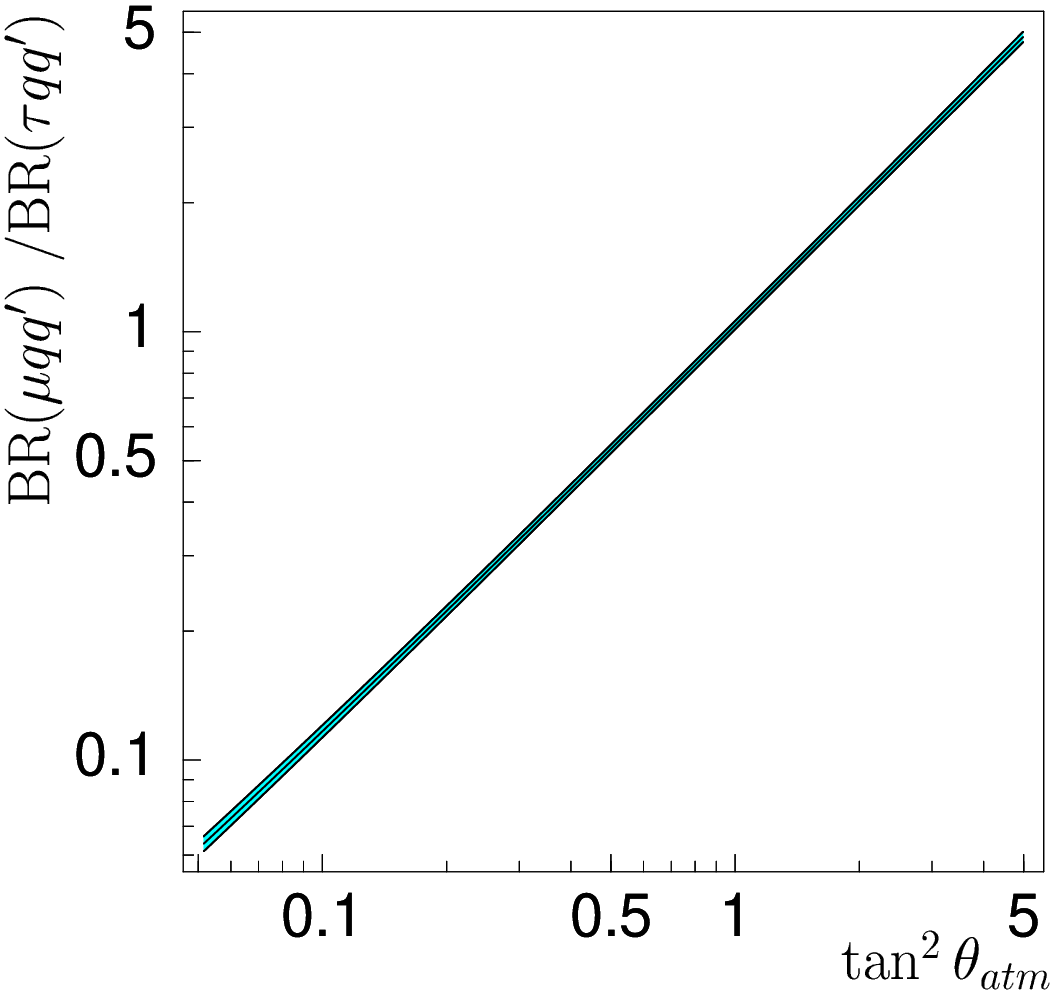}}
\end{picture}
&
\begin{picture}(7.0,6.5)
\put(0,0){\includegraphics[width=7cm]{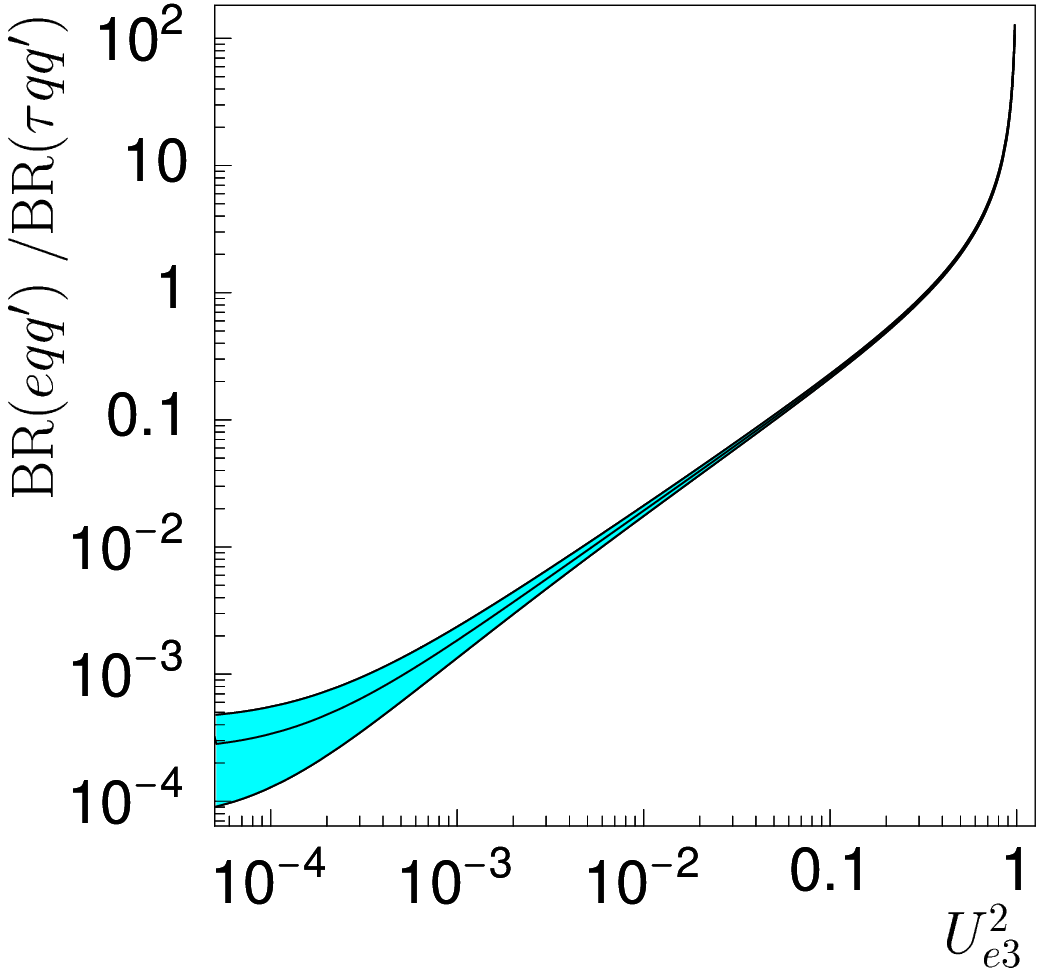}}
\end{picture}
\end{tabular}
\vspace{-3mm}
\caption{\small
The same as in Fig. 5 but for a unique SUSY point.
}
\label{corfu_fig6}
\end{figure}

If R-parity is broken, the neutralino is unstable and it will decay
through the following channels: $\chiz{1} \to \nu_i \, \nu_j \,
\nu_k,\ \nu_i \, q \, \bar{q}, \ \nu_i \, l^+_j \, l^-_k, \ l^\pm_i \,
q \, \bar{q}', \ \nu_i \, \gamma $. It was shown in
Ref.~\cite{Porod:2000hv}, that the neutralino decays well inside the
detectors and that the visible decay channels are quite large. This
was fully discussed in Ref.~\cite{Porod:2000hv} and is illustrated in
Fig.~\ref{corfu_fig4}.
We have seen before that the ratios $|\Lambda_i/\Lambda_j|$ and
$|\epsilon_i/\epsilon_j|$ were very important in the choice of solutions
for the neutrino mixing angles. What is exciting now, is that these
ratios can be measured in accelerator experiments.  In
Fig.~\ref{corfu_fig5} a) we show the ratio of branching ratios
for semileptonic LSP decays into muons and taus: $BR(\chi \to \mu q'
\bar q)/ BR(\chi \to \tau q' \bar q$) as function of $\tan^2
\theta_{atm}$. We can see that there is a strong correlation.  In
Fig.~\ref{corfu_fig5} b) is shown the ratio of branching ratios
for semileptonic LSP decays into muons and taus: $BR(\chi \to e q'
\bar q)/ BR(\chi \to \mu q' \bar q$) as function of $U_{e3}^2$. Again
we obtain a strong correlation. The spread on those figures can in
fact be explained by the fact that we do not know the SUSY
parameters. This is illustrated in Fig.~\ref{corfu_fig6} where
we considered that SUSY was already discovered with the following
values for the parameters,
\begin{equation}
M_2=120\ GeV,\ \mu=500\ GeV,\ \tan\beta=5,\
 m_0=500\ GeV,\ A=-500\ GeV
\end{equation}

\section{Conclusions}

The Bilinear R-Parity Violation Model is a simple extension of 
the MSSM that leads to a very rich phenomenology.
We have shown that the radiative breaking of the Gauge
Symmetry can be achieved in this BRpV Model.
We have calculated the one--loop corrected masses
and mixings for the neutrinos in a completely consistent way,
including the RG equations and correctly minimizing the potential.
We have shown that it is possible to get easily maximal mixing for the
atmospheric neutrinos and both small and large angle MSW.
We emphasize that the 
lightest neutralino decays inside the detectors,
thus leading to a very different phenomenology than the MSSM. 
If the model is to explain solar and atmospheric neutrino problems
many signals will arise at future colliders. These will probe
the neutrino mixing parameters. Thus the model is easily falsifiable!


\begin{thebibliography}{}  
\bibitem{kamiokande}
Y. Fukuda et al., Super-Kamiokande collaboration, {\it Phys.
Lett.} {\bf B433} (1998) 9; {\it Phys. Lett.} {\bf B436} (1998) 33; 
{\it Phys. Rev. Lett.} {\bf 81} (1998) 1562.

\bibitem{Gonzalez-Garcia}
M.~C.~Gonzalez-Garcia, M.~Maltoni, C.~Pena-Garay and J.~W.~F.~Valle,
Phys.\ Rev.\ D {\bf 63} (2001) 033005 [hep-ph/0009350];
%
N. Fornengo, M.C. Gonzalez-Garcia and J.~W.~F. Valle, Nucl. Phys. B {\bf
 580} (2000) 58;
%
M.~C.~Gonzalez-Garcia, H.~Nunokawa, O.~L.~Peres and J.~W.~F.~Valle,
Nucl.\ Phys.\ B {\bf 543} (1999) 3 [hep-ph/9807305].

\bibitem{chooz}  M. Apollonio et al., Chooz collaboration, {\it Phys. Lett.}
{\bf B420} (1998) 397.

\bibitem{BP98} 
J. N. Bahcall, S. Basu and M. H. Pinsonneault,  {\it Phys. Lett.} 
{\bf B 433}, (1998) 1.

\bibitem{Smy:2001yn}
M.~B.~Smy,
``Implications for solar neutrino oscillations from Super-Kamiokande
and  SNO data,'', To appear in the proceedings of NO-VE International
Workshop on Neutrino Oscillations in Venice, Venice 2001,
arXiv:hep-ex/0108053.


\bibitem{numass}
J. C. Rom\~ao, M. A. D\'{\i}az, M. Hirsch, W. Porod and J. W. F Valle,
{\it Phys. Rev.}{\bf D61}, (2000) 071703 ;
M. Hirsch, M. A. D\'{\i}az, W. Porod, J. C. Rom\~ao and J. W. F. Valle,
{\it Phys. Rev.}{\bf D62}, (2000) 113008.


\bibitem{epsmodel}
M. A. D\'{\i}az, J. C. Rom\~ao and  J. W. F. Valle {\it Nucl. Phys.}
{\bf 524}, (1998) 23.


\bibitem{marco}
M. A. D\'{\i}az (1998) Proceedings of the {\it International
Workshop on Physics Beyond the Standard Model: From Theory to
Experiment} (Valencia 97), Eds. Antoniadis, I., Iba\~nez, L.E. and
Valle, J.W.F., World Scientific, pag. 188. 


\bibitem{OriginalSBRP}
C. Aulakh, R. Mohapatra {\it Phys. Lett.} {\bf B119}, (1983)  136;
A. Santamaria, J. W. F. Valle {\it Phys. Lett.} {\bf B195}, (1987) 423;
A. Santamaria and J. W. F. Valle  {\it Phys. Rev. Lett.} {\bf 60},
(1988) 397.

\bibitem{SBRpV}
A. Masiero and J. W. F. Valle {\it Phys. Lett.} {\bf B251}, (1990) 273;
J. C. Rom\~ao, C. A. Santos and J. W. F. Valle {\it Phys. Lett}
{\bf B288}, (1992) 311.



\bibitem{HPVR2002}
M. Hirsch, M. A. Di\'az, W. Porod, J. W. F. Valle and J. C. Rom\~ao, work in
preparation.  


\bibitem{Porod:2000hv}
W.~Porod, M.~Hirsch, J.~Rom\~ao and J.~W.~Valle,
{\it Phys.\ Rev.}\ {\bf D63} (2001) 115004.



\end{thebibliography}
\end{document}